\documentclass[screen, acmsmall]{acmart}
\pdfoutput=1
\usepackage[utf8]{inputenx}
\usepackage[disable]{todonotes}
\usepackage{booktabs}
\makeatletter
\def\mdseries@tt{m}
\makeatother
\usepackage[frozencache]{minted}
\usepackage{tikz}
\usetikzlibrary{positioning, shapes, arrows.meta, calc, fit}
\usepackage[margin=0.3em]{subcaption}

\makeatletter
\tikzset{
    database/.style={
        path picture={
            \draw (0, 1.5*\database@segmentheight) circle [x radius=\database@radius,y radius=\database@aspectratio*\database@radius];
            \draw (-\database@radius, 0.5*\database@segmentheight) arc [start angle=180,end angle=360,x radius=\database@radius, y radius=\database@aspectratio*\database@radius];
            \draw (-\database@radius,-0.5*\database@segmentheight) arc [start angle=180,end angle=360,x radius=\database@radius, y radius=\database@aspectratio*\database@radius];
            \draw (-\database@radius,1.5*\database@segmentheight) -- ++(0,-3*\database@segmentheight) arc [start angle=180,end angle=360,x radius=\database@radius, y radius=\database@aspectratio*\database@radius] -- ++(0,3*\database@segmentheight);
        },
        minimum width=2*\database@radius + \pgflinewidth,
        minimum height=3*\database@segmentheight + 2*\database@aspectratio*\database@radius + \pgflinewidth,
    },
    database segment height/.store in=\database@segmentheight,
    database radius/.store in=\database@radius,
    database aspect ratio/.store in=\database@aspectratio,
    database segment height=0.1cm,
    database radius=0.25cm,
    database aspect ratio=0.35,
}
\makeatother

\excludecomment{outline}

\graphicspath{{figures/}}
\newcommand{\omax}{$\omega_{\text{max}}$}

\begin{document}

\title{Optimally Self-Healing IoT Choreographies} 

\author{Jan Seeger}
\email{jan.seeger@siemens.com}
\orcid{0000-0001-6547-2306} 
\affiliation{%
  \institution{TU München}
  \streetaddress{Arcisstraße 21}
  \city{München}
  \postcode{80333}
  \country{Germany}}


\author{Arne Bröring}
\affiliation{%
  \institution{Siemens AG}
  \streetaddress{Otto-Hahn-Ring 6}
  \city{München}
  \postcode{81739}
  \country{Germany}}

\author{Georg Carle}
\affiliation{%
  \institution{TU München}
  \streetaddress{Arcisstraße 21}
  \city{München}
  \postcode{80333}
  \country{Germany}}


\begin{abstract}
  In the industrial Internet of Things domain, applications are moving from the Cloud into the edge, closer to the devices producing and consuming data.
  This means applications move from the scalable and homogeneous cloud environment into a constrained heterogeneous edge network.
  Making edge applications reliable enough to fulfill Industrie 4.0 use cases is still an open research challenge.
  Maintaining operation of an edge system requires advanced management techniques to mitigate the failure of devices.
  This paper tackles this challenge with a twofold approach: (1) a policy-enabled failure detector that enables adaptable failure detection and (2) an allocation component for the efficient selection of failure mitigation actions.
  We evaluate the parameters and performance of our failure detection approach and the performance of an energy-efficient allocation technique, and present a vision for a complete system as well as an example use case.
\end{abstract}



\maketitle

\section{Introduction}

IoT application deployments are currently mostly cloud-based, with a central processing component provided with data via remote sensors and actuators.
This centralized cloud structure limits the possible applications for latency and confidentiality reasons.
As a response, processing is moving back into the edge of the network, or into the sensors and actuators themselves.
Centralized cloud systems are being supplanted by edge systems in latency- and privacy-sensitive applications.
Industrial applications have stringent requirements in these fields, with low latency necessary for monitoring and control applications, and confidentiality necessary for economic reasons.

Shi et al.~\cite{shi_edge_2016} define the term ``edge'' as ``any computing and network resources along the path between data sources and cloud data centers''.
We use this definition, but focus on edge networks that are close to the data producers and consumers and fully under the control of the application owner.
Thereby, applications consist of multiple chained tasks, which can be distributed over several edge nodes to enable collaboration, e.g., to process a complex algorithm or AI pipeline in a distributed and coordinated way.
Challenge is the heterogeneity of industrial edge networks, which consist of various kinds of nodes (e.g., industrial PCs, HMI units, network switches, or field devices such as simple sensors and actuators).
These nodes have varying computational and communication capacities.
Hence, operating a distributed application reliably and efficiently requires intelligent management approaches.

In this paper, we focus on (1) efficiently detecting failures of devices and software components using an accrual-based failure detection augmented with policies, and (2) automatically mitigating failures by finding an optimal allocation of application tasks, e.g., towards minimized energy consumption of the system.
This work describes the latest findings on our research agenda to enable distributed IoT choreographies.
Our path began with the introduction of the ``Recipe'' concept for defining IoT application templates~\cite{thuluva_recipes_2017}, continued by our work on improving the runtime management of such recipes by handling them as service choreographies~\cite{seeger_running_2018}, and most recently defined a mechanism for the dynamic and resilient management of IoT choreographies~\cite{seeger_dynamic_2019}.

A ``Recipe'' describes an application template as a graph of abstract IoT tasks (e.g., device services such as ``stream video'' or ``notify user'') where data flows along the edges of the graph, and tasks are executed when they have received enough input data.
An example recipe use case implementing a vibration analysis for rotating machinery is described in more detail in Section~\ref{sec:system-model}.

Recipe tasks are generic, and a concrete recipe is derived by ``instantiating'' these generic tasks with concrete implementations that are available in the system.
Using this mechanism, parts of the application can be replaced automatically when failure is detected.
Missing in our previous work are evaluations of the ``quality'' of these replacements.
So far, the first replacement available was chosen without regarding the effect on the properties of the system.
Such relevant properties are for example the end-to-end latency of the application or the energy usage.
By optimizing the placement of tasks on devices, we can optimize the properties of a system even when devices fail.
Building up on our previous work, we present here a thorough examination of our failure detector and an evaluation of its memory and processing usage, as well as a detailed description and evaluation of our mechanism to optimize the assignment of tasks to devices for minimal energy usage.

\section{Background \& Related Work}\label{sec:background--related}

In this section, we describe the context of our work.
Section~\ref{sec:iot-composition} introduces relevant works in the field of composing services (and IoT devices) to applications.
Section~\ref{sec:failure-detection} provides an overview about mechanisms for detecting failures in distributed systems.
Section~\ref{sec:optimal-allocation} presents related work on optimal placements of system operators.

\subsection{IoT Composition}\label{sec:iot-composition}

The composition of web services has been extensively researched~\cite{sheng_web_2014}.
Thereby, service composition can be classified into two types, service orchestration and service choreography, based on the manner in which the participant services interact~\cite{sheng_web_2014}.
Productive solutions for web services composition typically follow the orchestration approach.

In the IoT domain, we have today established composition systems with a broad user community such as ``If This Then That''\footnote{\url{http://ifttt.com}} and Node-RED\footnote{\url{http://nodered.org}}.
These tools use simple composition techniques that are executed centrally as orchestrations.
These platforms are targeting mainstream users and lack systematic engineering support, which leads e.g.\ to widely duplicated recipes, as shown by Ur\cite{ur_trigger-action_2016}.

Giang et al.~\cite{giang_developing_2015} focus on application-level distributed choreographies by building on Node-RED as a visual programming tool.
However, they do not address the configuration of critical automation systems and their need for failure detection and recovery.

Khan et al.\cite{khan_reliable_2017} propose a reliable infrastructure for IoT compositions, but focus on communication of data instead of application-level orchestrations.
Thuluva et al.\cite{thuluva_semantic-based_2017} employ Semantic Web technologies to enable low-effort engineering of industrial IoT applications.
However, they do not focus on the runtime aspects and dynamic reconfiguration or failure detection.
Focusing on building automation, Ruta et al.\cite{ruta_semantic-based_2014} present a multi-agent framework that uses semantic technologies and makes use of automated reasoning for enabling device discovery and orchestration of IoT components.
Their approach misses to address failure handling or mitigation.

This work builds up on our previous works~\cite{thuluva_recipes_2017, seeger_dynamic_2019,seeger_running_2018} that present an IoT composition as a ``Recipe'', i.e., separate from its implementation.
A semi-automated service composition and instantiation tool assists the user in creating the composition.
IoT choreographies are described by a directed graph of connected abstract application components, with data flowing along the edges of these components.
During the instantiation phase, each abstract component is replaced with a concrete one.
The approach of this work builds up on this concept by rerunning the instantiation algorithm when a failure is detected, to find another component that can fulfill the functionality of the failed component.

\begin{figure}
  \centering
  \includegraphics[scale=0.5]{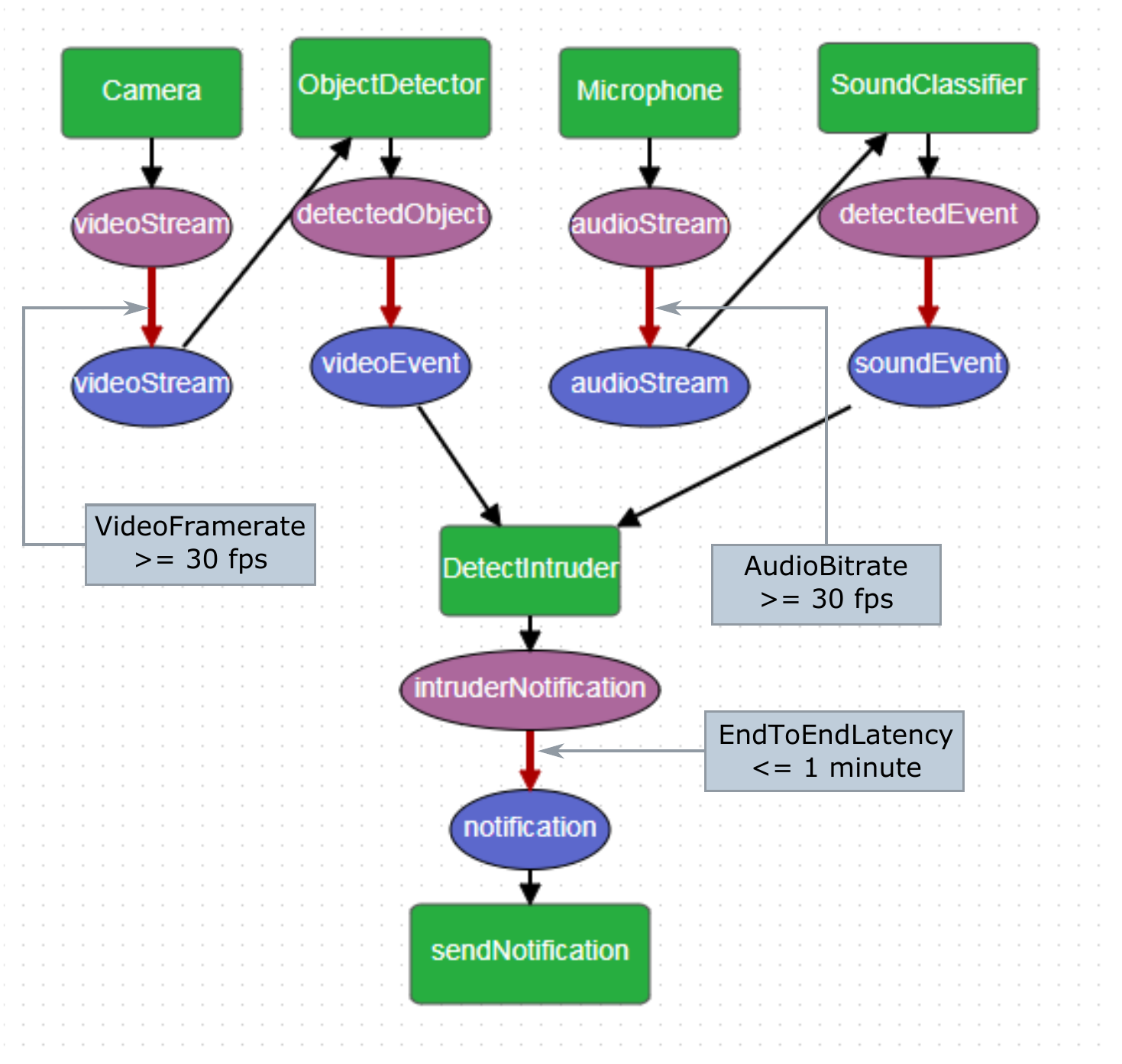} 
  \caption{\label{fig:example-recipe} Example of a recipe combining multiple device services for object detection (Source:~\cite{Seeger_rule_based2019}).}
\end{figure}

Figure~\ref{fig:example-recipe} shows an example of a recipe that combines multiple services of devices in an intrusion detection system. The green boxes are ingredients that need to be replaced with concrete components when the recipe is instantiated. Ingredients are connected via their outputs and inputs. 
In this example, video and audio streams are connected to analytics components that feed into an aggregating intrusion detector and finally a component that is able to send notifications. The recipe designer can further specify application-level constraints (e.g., minimum video frame rate) on the interactions between sensors and analysis services~\cite{Seeger_rule_based2019}.

\subsection{Failure Detection}\label{sec:failure-detection}

Failure detection is an essential building block for distributed systems.
Without a suitable failure detector, distributed applications are generally not guaranteed to complete successfully.
Chandra et al.\ describe the theory of failure detection in~\cite{chandra96_unreliable_failure_detectors_reliable_distributed_systems} and define two properties of failure detectors: \emph{completeness} and \emph{accuracy}.
Completeness describes the property of a failure detector to correctly detect failure, while accuracy describes the capability of a detector to not detect failure on nodes functioning correctly.

Our research is based on $\phi$-accrual failure detection, which is in detail described by Défago et al.~\cite{defago04_phi_accrual_failure_detector} and more recent implementations include Satzger et al.~\cite{satzger07new_adaptive_accrual_failure_detector} and Liu et
al.~\cite{liu_energy-efficient_2018}. Accrual failure detectors calculate the probability of a node having failed from the distribution of inter-arrival times of received failure detection messages.
Thereby, accrual failure detectors are strongly \emph{complete} (there is some time after which all failed processes are permanently marked failed by all other processes) and eventually strongly \emph{accurate} (there is some time after which correct processes are not marked failed by other correct processes).


In IoT environments, failures have to be detected on all involved devices and services, in order to ensure reliability. Kodeswaran et al.~\cite{Kodeswaran2016} present a system for efficient failure management in smart home environments based on tracking the most performed activities. This knowledge is used to predict future degradation and failures of involved devices. However, their work is restricted to the smart home domain and not directly usable in general IoT or industrial IoT cases.

The Gaia framework~\cite{Chetan2005} allows to build pervasive systems and also includes failure detection. Thereby, the implemented failure detector is part of a central controller. I.e., the availability of the controller has to be ensured for the failure detection to work. Our approach is based on the distributed nodes and does not require communication with a central controller to detect a failure. 

In~\cite{de15new_unreliable_failure_detector_for_self_healing_in_ubiquitous_environments} an unreliable failure detector is presented that enables the definition of an impact factor for the involved nodes.
This allows to tune the performance of the failure detector for specific application needs.

Guclu et al.~\cite{guclu_distributed_2016} present a distributed failure detector that builds on trust management.
Their method evaluates the trustworthiness of the data from neighboring nodes.
However, their approach is limited to homogeneous networks and similar structured data.
This makes it not applicable to our scenarios of heterogeneous edge and industrial IoT environments.

\subsection{Optimal Allocation}\label{sec:optimal-allocation}

Efficiently allocating application \emph{tasks} of a recipe to available edge devices is comparable to a widely studied research problem in the distributed systems field: the optimal \emph{operator} placement of distributed stream processing applications, or the optimal selection of networked devices for \emph{tasks} or \emph{computations} of a chained process or workflow.
When either a hardware or software failure occurs, an application \emph{component} has failed, and this failure needs to be mitigated.

Tasks have different parameters, depending on the optimization target.
The result of the task allocation problem is an allocation, an assignment of tasks to devices, that fulfills the constraints, and improves the performance of the system in some metric.

An overview of existing allocation approaches for stream processing is given in~\cite{lakshmanan_placement_2008}.

Based on Constraint Programming, Haubenwaller \& Vandikas~\cite{Haubenwaller2015} describe an approach for the efficient distribution of actors (processing tasks) to IoT devices.
The approach resembles the Quadratic Assignment Problem and is NP-hard, resulting in long computation times when scaling up.
Samie et al.~\cite{Samie2016} present another Constraint Programming-based approach that takes into account the bandwidth limitations and minimizing energy concumption of IoT nodes.
The system optimizes computation offloading from an IoT node to a gateway, however, it does not consider composed computations that can be distributed to multiple devices.

A Game Theory-based approach is presented in~\cite{Sardellitti2015} that aims at the joint optimization of radio and computational resources of mobile devices.
The system local optimum for multiple users, however, it only aims at deciding whether to fully offload a computation or to fully process it on device.


Based on Non-linear Integer Programming, Sahni et al.~\cite{Sahni2017} present their Edge Mesh algorithm for task allocation that optimizes overall energy consumption and considers data distribution, task dependency, embedded device constraints, and device heterogeneity.
However, only basic evaluation and experimentation are done and no performance comparison has been performed.

Based on Integer Linear Programming (ILP), Mohan \& Kangasharju~\cite{Mohan2016} propose a task assignment solver that first minimizes the processing cost and secondly optimizes the network cost, which stems from the assumption that Edge resources may not be highly processing-capable.
An intermediary step of reduces the sub-problem space by combining tasks and jobs with the same associated costs.
This reduces the overall processing costs.


Cardellini et al.~\cite{cardellini16optimal_operator_placement} describe a comprehensive ILP-based framework for optimally placing operators of distributed stream processing applications, while being flexible enough to be adjusted to other application contexts.
Different optimization goals are considered, e.g., application response time and availability.
They propose their solution as a unified general formulation of the optimal placement problem and provide a strong theoretical foundation.
The framework is flexible so that it can be extended by adding further constraints or shifted to other optimization targets.
Hence, we utilize their framework and extend it by incorporating further constraints for our optimization goal, the overall energy usage.

\section{A Failure Detector for Self-Healing IoT Choreographies}\label{sec:fail-detect-with}

In this section, we describe our failure detector PE-FD and its properties in comparison to other failure detectors, and present our policy concept to tune the failure detector for specific application requirements.

\begin{outline} Motivation for failure detector.
  Go into failure detector theory (\cite{chandra96_unreliable_failure_detectors_reliable_distributed_systems}, describe accrual detector from~\cite{defago04_phi_accrual_failure_detector}).
\end{outline}

\subsection{The PE-FD Failure Detector}\label{sec:iota-fail-detect}

\begin{outline} Describe implementation.
  Reference~\cite{seeger_dynamic_2019}, but describe in more detail (go into details on Memory consumption and complexity of suspicion update).
  Compare with two other approaches (\cite{defago04_phi_accrual_failure_detector, satzger07new_adaptive_accrual_failure_detector})
\end{outline}

Failure detection is a crucial functionality for a distributed system that should operate reliably.
Without the information on the current state of components, a mitigator cannot decide what (if any) action to take to continue operation of the system.
In this work, we focus on crash failures~\cite{tanenbaum_distributed_2007}, where devices and software work correctly until they fail permanently.
For efficient failure detection in IoT Choreographies, we have developed our Policy-Enabled Failure Detector (PE-FD).
It is based on the principle of $\phi$-accrual failure detection~\cite{defago04_phi_accrual_failure_detector} and augmented with the support for ``policies'', where parameters of the failure detection algorithm are adjusted according to application requirements.

In general, $\phi$-accrual failure detectors are unreliable, meaning that errors in their output are permissible, but after some point, their output is always correct.
Compared to ``traditional'' failure detectors (e.g., heartbeat-based or adaptive~\cite{bertier_implementation_2002, chen_quality_2002}), $\phi$-accrual detectors compute a suspicion function $\phi$ that describes the probability of a node having crashed.
This probability is computed by estimating the distribution of the inter-arrival times of heartbeats, and computing the probability of a new heartbeat arriving after the current time.
With the current time being $t_{\mathit{now}}$, and the time of the last timestamp's arrival being $t_{\mathit{last}}$, the suspicion is thus given by the formula $\phi(t_{\mathit{now}}) \equiv -\log_{10}(P_{\mathit{later}}(t_{\mathit{now}} - t_{\mathit{last}}))$, where $P_{\mathit{later}}$ is computed from the inferred distribution of timestamp inter-arrival times.

Failure detectors differ in the parameters and implementation of estimating $P_{\mathit{later}}$, either storing all inter-arrival times for every heartbeat and using the empirical distribution function, or assuming a certain distribution for inter-arrival times and estimating the parameters of the distribution.
Our PE-FD failure detector computes suspicion in constant time and space by taking advantage of the one-sided Chebyshev inequality together with empirical estimators for both $\phi$ and $\mu$.

\begin{align}
  \rho_n &= \sum_{i=1}^n x_i = \rho_{n-1} + x_i\label{eq:1}\\
  \kappa_n &= \sum_{i=1}^{n} x_i^2 = \kappa_{n-1} + x_n^2\label{eq:2}\\
  \mu_n &= \frac{1}{n} \rho_n\label{eq:3}\\
  \sigma^2_n &= \frac{1}{n-1}\sum_{i=1}^{n}{(X_i - \mu)}^2 = \frac{1}{n-1}\left[\kappa_n - n\mu_n^2\right]\label{eq:4}\\
  P_{\mathit{later}} &= P[X > \mathit{now}] \leq \frac{\sigma^2}{\sigma^2 + {\left(T_{\mathit{now}} - \mu\right)}^2}\label{eq:5} \\
  P_{\mathit{later}} &= 0 | \mathit{now} < mu\label{eq:15}
\end{align}

In equations~\ref{eq:1} and~(\ref{eq:2}), we define two helper variables that store the sum of the timestamps and the sum of the squares of the timestamps.
We can then compute the mean $\mu$ by dividing by the number of timestamps, and derive $\sigma^2$ via equation~(\ref{eq:4}).
To calculate mean and variance, we thus need to store three variables ($\rho$, $\kappa$ and $n$), independent of the number of timestamps received.
The suspicion can then be calculated without any additional information.
To combat overflow and numerical instability, $\rho$ and $\kappa$ are periodically reset after a number of timestamps have been received.
We call this number $\omega_{\text{max}}$ ``learning window''.
However, after resetting, we need a certain number of heartbeats to regain a good estimate for the distribution parameters.
Thus, we introduce a parameter $\omega_{\min}$ that is the minimum number that needs to be received until the new estimate is used.

\begin{listing}
  \centering
\begin{minted}{python}
def updateState(state, delay):
    return IotaState(state.rho + delay, state.kappa + delay**2, state.n + 1)

def suspicion(state, time):
    if state.n == 0 or state.n = 1:
        return 0
    mu = state.rho / state.n
    if time < mu:
        return 0
    sigma = (state.kappa - state.n * (mu * mu)) / (state.n - 1)
    return -np.log10(sigma / (sigma + (time - mu)**2))
\end{minted}
  \caption{\label{lst:iota-fail-detect}Suspicion calculation and state update for the PE-FD failure detector.}
\end{listing}

A python implementation of the suspicion algorithm is included in Listing~\ref{lst:iota-fail-detect}.
It can be seen that the calculation of the next state takes a constant amount of computations (three additions, and one multiplication).
The necessary operations to calculate a suspicion value are three divisions, three multiplications and four additions.
The number of computations is independent of the chosen parameters.
With extremely large values for the learning window, the variables $\rho$ and $\kappa$ could overflow, but this is not a realistic constraint for 32-bit variables.

$\Phi$-accrual failure detectors can be converted into binary failure detectors by choosing a threshold $u$, and marking a process as failed when the suspicion rises above this threshold.
We describe the choices for $u$ in the next section, and the behavior of our PE-FD failure detector is evaluated in detail in Section~\ref{sec:evaluation-iota-fd}.

\subsection{Policies for application-tuned failure detection}\label{sec:polic-appl-tuned}

The PE-FD failure detector provides a number of adjustable parameters:

\begin{itemize}
\item The minimum number of heartbeats required for an estimate ($\omega_{\min}$)
\item The maximum number of heartbeats until the current estimate is reset  ($\omega_{\max}$)
\item The heartbeat period ($t_{\text{heartbeat}}$)
\item The suspicion threshold ($u$)
\end{itemize}

$\omega_{\min}$ and $\omega_{\max}$ are called the ``learning window'' in combination.
These parameters represent a wide range of adaptability for our algorithm.
By adjusting these parameters based on policies that take the structure and requirements of applications into account, failure detection can be improved over a ``one-size-fits-all'' approach.

The maximum and minimum number of heartbeats $\omega_{\min}$ and $\omega_{\max}$ are relevant for nodes with changing network conditions.
For example, a mobile node can benefit from a lower maximum and minimum number of heartbeats, so the failure detection algorithm can adapt to changing network conditions with fewer received heartbeats.
The complementary adjustment is possible as well: For wired nodes, increasing the size of the learning window allows them to ignore transient failures, and keep the application working.

When failure detection is used in a web service composition as those described in Section~\ref{sec:iot-composition}, the structure of the composition can be inspected to modify the suspicion threshold and the heartbeat period.
When a task is central to an application, and no replacement is available, the heartbeat period should be set low to allow quick detection of failures.
The suspicion threshold $u$ should be set relatively high, not to cause false positives.



\section{Optimal mitigation of failures in IoT Choreographies}
\label{sec:optimal-mitigation}

In this section, we describe how our failure detector is combined with a task assignment approach to an optimal self-healing procedure for IoT choreographies, and we describe the details of allocating tasks to optimize the overall energy usage.

\subsection{Optimal Self-Healing Procedure}

IoT applications are becoming more prevalent in various domains, e.g., smart homes and buildings, industrial manufacturing, transportation, or healthcare.
Such IoT applications often consist of multiple tasks that interact in the form of a dataflow graph, where components exchange data along directed edges.
A simple but popular execution engine for such application is ``IFTTT'' (Section~\ref{sec:iot-composition}).

With the recipe concept (Section~\ref{sec:iot-composition},~\cite{thuluva_recipes_2017}), we have defined a schema for the expression of such dataflow graphs.
These recipes are executed in a distributed fashion, and can be dynamically replaced and reconfigured (see~\cite{seeger_running_2018}).
With the growing scale of IoT device deployments and applications, such dynamic replacement and reconfiguration will become more important.
Additionally, IoT applications are penetrating more an more crucial areas (e.g., patient monitoring or optimization of industrial processes), i.e., failures may have large impact and the ability to reconfigure the system dynamically is crucial.

In our previous work, the chosen replacement component was not evaluated for its quality, besides fulfilling the obvious functional requirements.
Hence, we aim here at evaluating the replacement of devices with regard to a defined quality metric and thereby  improve the operating parameters of the application and the network.
This is especially important with long-running and resource-constrained processes.

Combined with the failure detection presented in Section~\ref{sec:failure-detection}, the steps of our optimal self-healing procedure for an IoT orchestration are as follows:

\begin{enumerate}
\item Detect software or device failure
\item Find functionally matching replacement devices 
\item Optimal assignment of application tasks to available network nodes
\item Reconfigure application with replacement device and software
\end{enumerate}

Everything in that application is allocated.
We can efficiently detect software or device failure (1) via the failure detection algorithm described in Section~\ref{sec:fail-detect-with}, and tune the detection according to the application requirements with policies as described in Section~\ref{sec:polic-appl-tuned}.
The semantic matching algorithm in~\cite{seeger_running_2018,thuluva_recipes_2017} can then find a functionally matching replacement component (2).
Then, in step (3), we evaluate the placement of recipe components on nodes of the modified graph via the allocation algorithm described in the next Section~\ref{sec:energy-optimal-task}.
Step (4), the reconfiguration of the instantiated recipe, then happens as described in~\cite{seeger_running_2018}.

\subsection{Energy-optimal Task Assignment}\label{sec:energy-optimal-task}

\begin{outline} Describe response time optimization from~\cite{cardellini16optimal_operator_placement}, describe our
  energy optimization approach.
\end{outline}

As described in Section~\ref{sec:optimal-allocation}, we have based our approach for allocating recipe tasks on~\cite{cardellini16optimal_operator_placement}.
Cardellini et al.\ evaluate configuring a system for optimal response time and optimal availability.
They formulate the allocation of operators as an ILP problem, which they hand over to an IBM CPLEX\footnote{\url{https://www.ibm.com/analytics/cplex-optimizer}} solver to find the optimal approach.
We have followed this approach to optimize the energy usage of an IoT application formulated as a recipe.

We define optimality of the allocation by total energy use over one execution of the recipe.
Energy during recipe execution is consumed in two phases: ``Device energy'' is consumed by a device when executing a task, and ``network energy'' is consumed by the device when sending the result of the calculation over the network.
The optimal configuration of the network is the assignment of tasks to devices that results in the lowest total consumption of energy and satisfies the constraints.
The constraints concern the requirements that an assignment must satisfy: Each task should only be allocated once and resource requirements for assigned tasks should not exceed the resources of the node.
This problem is a form of the quadratic assignment problem, and thus NP-hard.
We have developed and evaluated a heuristic that reduces the problem to a non-quadratic assignment problem, which we describe in Section~\ref{sec:optim-energy-optim}.

We define the energy-optimal task assignment as follows: The recipe $G_{\text{rcp}}$consists of a set of tasks $V_{\text{rcp}}$ connected by directed edges $E_{\text{rcp}}$.
The network $G_{\text{net}}$ that tasks can be evaluated on consists of a set of nodes $G_{\text{net}}$ connected by a set of undirected links $E_{\text{net}}$.
The result of the allocation is a matrix $X = V_{\text{rcp}} \times V_{\text{net}}$ where $X[t, n] = 1$ if and only if task $t$ is allocated to node $n$.

\begin{table}
  \centering
  \begin{tabular}{ll}
    \toprule
    Symbol & Description\\
    \midrule
    $R_t$ & Resources required for the evaluation of task $t \in V_{\text{rcp}}$.\\
    $O_t$ & Output of task $t \in V_{\text{rcp}}$ for a single received input.\\
    $S_t$ & Computation time required for completing task $t \in V_{\text{rcp}}$ once.\\
    $P_n$ & Processing power of node $n \in V_{\text{net}}$.\\
    $R_n$ & Resources available on node $n \in V_{\text{net}}$.\\
    $C_n$ & Energy consumption of node $n \in V_{\text{net}}$ for one unit of computation.\\
    $T_l$ & Energy use for the transfer of one data packet over link $l \in E_{\text{net}}$.\\
    $D_{(n_1, n_2)}$ & Energy cost of the shortest path between $n_1$ and $n_2$.\\
    \bottomrule
  \end{tabular}
  \caption{\label{tab:allocation-parameters}Parameters of energy-aware allocation algorithm.}
\end{table}

Tasks, nodes and links have properties that are relevant for the energy consumption of the application once allocated.
These parameters are described in Table~\ref{tab:allocation-parameters}.
$S_t$, $P_n$, $R_n$ and $C_n$ are defined as multiples of some reference node.
The resources of a node are expressed as a single scalar, but additional resource requirements can easily be introduced into the model.

For calculating the network energy, we need to know whether a link between two tasks is assigned to a link between two nodes.
For this, we introduce a matrix $Y= V_{\text{rcp}} \times V_{\text{rcp}} \times V_{\text{net}} \times V_{\text{net}}$, where $Y[t_1, t_2, n_1, n_2] = 1$ if and only if the communication between task $t_1$ and task $t_2$ is allocated on the network link between nodes $n_1$ and $n_2$.
This corresponds to $X[t_1, n_1] = 1 \wedge X[t_2, n_2]$.
Unfortunately, this is not a linear constraint, and thus we need to linearize the formulation.

\begin{align}
  \forall t_1, t_2 \in V_{\text{rcp}}: \forall n_1, n_2 \in V_{\text{net}}: Y[t_1, t_2, n_1, n_2] &\le X[t_1, n_1]\label{eq:10}\\
  \forall t_1, t_2 \in V_{\text{rcp}}: \forall n_1, n_2 \in V_{\text{net}}: Y[t_1, t_2, n_1, n_2] &\le X[t_2, n_2]\label{eq:11}\\
  \forall t_1, t_2 \in V_{\text{rcp}}: \forall n_1, n_2 \in V_{\text{net}}: Y[t_1, t_2, n_1, n_2] &\ge X[t_1, n_1] + X[t_2, n_2] - 1\label{eq:12}\\
  \forall t \in V_{\text{rcp}}: \sum_{n \in V_{\text{net}}} X[t,n] &= 1\label{eq:6}\\
  \forall n \in V_{\text{net}}: \sum_{t \in V_{\text{rcp}}} X[t,n] * R_t &\le R_n\label{eq:7}\\
  \sum_{t \in V_{\text{rcp}}}\sum_{n in V_{\text{net}}} C_n * (S_t / P_n) * X[t,n] &\le \text{device\_energy}\label{eq:8}\\
  \sum_{(t_1, t_2) \in E_{\text{rcp}}} \sum_{n_1, n_2 \in V_{\text{net}}} O_{n_1} * P_{n_1,n_2} * Y[t_1, t_2, n_1, n_2] &\le \text{network\_energy}\label{eq:9}\\
  \text{network\_energy} + \text{device\_energy} &\le \text{total\_energy}\label{eq:14}
\end{align}

We follow the formulation presented in~\cite{cardellini16optimal_operator_placement} and define an ILP model as shown in Equations~\ref{eq:10} to~\ref{eq:12}.
Equations~\ref{eq:10} to~\ref{eq:12} describe the linearization of the network matrix $Y$.
Equations~\ref{eq:6} and~\ref{eq:7} express the ``only allocated once'' and ``resources not exceeded'' constraint.
Equations~\ref{eq:8} and~\ref{eq:9} calculate network and device energy as described above.
Finally, we calculate the total energy use of the assignment by adding both energies in equation~\ref{eq:14}.
The objective of the optimization is the minimization of the total used energy.

We implemented this model in a Python\footnote{\url{https://www.python.org/}} script using the PuLP\footnote{\url{https://pythonhosted.org/PuLP/}} linear programming library.
We can then find solution for the problem using the CPLEX solver, which uses a branch-and-bound approach \cite{ross_branch_1975}.
For a discussion of the benchmark results, see Section~\ref{sec:evaluation-mitigator}.

\subsection{A Linear Heuristic for Energy-Optimized Allocation }\label{sec:optim-energy-optim}

The quadratic assignment problem described in the previous section is NP-hard and thus compute intensive.
The culprit for this is the network cost calculation and the linearization of Y resulting in a large number of constraints.
By approximating the network energy, we can get a faster solution, which is however no longer optimal.
We quantify the loss of optimality and speedup in Section~\ref{sec:evaluation-mitigator}.

By removing the $Y$ matrix and the associated constraints, we create a linear problem that can be evaluated effectively by the simplex method \cite{nelder_simplex_1965}.
Our approach approximates the energy required for sending a packet of data by taking the average of a node's links.
We introduce the parameter $\hat{T}_n = \frac{1}{|\text{outgoing}(n)|} \sum_{e \in \text{outgoing}(n)} T_e$ that describes the average transmission cost of a node's links.

\begin{align}
  \sum_{t \in V_{\text{rcp}}}\sum_{n in V_{\text{net}}} C_n * (S_t / P_n) * X[t,n] + O_{t} * \hat{T}_n * X[t,n] &\le \text{total\_energy}\label{eq:13}
\end{align}

The complete model reuses constraints~(\ref{eq:6}) and~(\ref{eq:7}) with the constraint (\ref{eq:13}).
By transforming the QAP into a linear problem, we greatly increase the speed of finding a solution, and make the optimization feasible for on-line usage.

\section{System Model \& Application Example}\label{sec:system-model}

This section presents our implementation of the system for optimal self-healing approach for IoT choreographies and describes a use case example for applying the developed system.

\begin{figure}
  \centering
  \begin{tikzpicture}[component/.style={draw, thick, minimum width=1cm, minimum height=1cm, align=center}, every edge/.style={draw, thick, >=Latex[], shorten >=.5mm, shorten <=.5mm}, node distance=1cm and 2cm]
    \node[component] (controller) {Configurator};
    \node[right=of controller, component, database, database radius=.7cm, database segment height=2mm, draw=none, label={[align=center]below:Knowledge\\base}] (kb) {};
    \node[component, left=of controller] (allocator) {Allocator};

    \node[below=2cm of controller] (dev2) {\includegraphics[width=.08\textwidth]{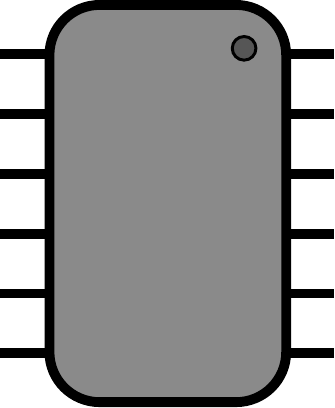}};
    \node[left=of dev2] (dev1) {\includegraphics[width=.08\textwidth]{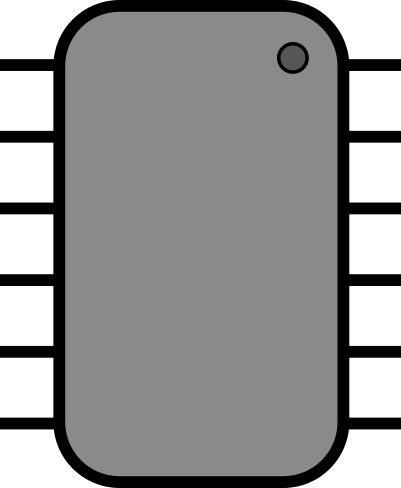}};
    \node[right=of dev2] (dev3) {\includegraphics[width=.08\textwidth]{micro}};

    \draw (dev2) edge [<->, double] node[auto,swap] {PE-FD} (dev1) edge [<->, double] node[auto] {PE-FD} (dev3);
    
    \node[component, ellipse, fit=(dev1) (dev2) (dev3),label=200:Network] (network) {};
    \path (controller) edge [<->]  node[auto] {(2)} (kb);
    \path ($(controller.west) + (0,1mm)$) edge [->] node[auto, yshift=-2mm] {(4)} ($(allocator.east) + (0,1mm)$);
    \path ($(allocator.east) - (0,1mm)$) edge [->] node[auto, yshift=2mm] {(3)} ($(controller.west) - (0,1mm)$);
    \path ($(controller.south) - (1mm,0)$) edge [->] node[auto, swap] {(5)} ($(network.north) - (1mm,0)$);
    \path ($(network.north) + (1mm,0)$) edge [->] node[auto, swap] {(1)} ($(controller.south) + (1mm,0)$);

  \end{tikzpicture}
  \caption{\label{fig:system-model-an} System model of an optimally self-healing IoT choreography}
\end{figure}

Figure~\ref{fig:system-model-an} shows the integrated system combining our optimal allocator and the PE-FD failure detector.
The system consists of 3 main components, as well as the devices in the network.
A \emph{knowledge base} stores the knowledge about the system, such as available devices, applications and links between devices.
One possible storage mechanism for such data would be a semantic triple store such as Apache Jena\footnote{\url{https://jena.apache.org/}}.
With this semantic store, the system can take advantage of semantic reasoning and translation, as described in~\cite{Seeger_rule_based2019}.
The \emph{configurator} controls the creation of the system and is responsible for configuring devices into a choreography.
It is not involved in the operation of the system, but for administrative actions (such as reconfiguring applications when devices fail).
The \emph{allocator} is the component responsible for running the allocation algorithms described in Section~\ref{sec:optimal-mitigation}.
Finally, the \emph{network} contains devices that communicate via heterogeneous network links.
These devices are running an engine that supports configuration by the configurator.

When a device or software component fails, this is detected by the PE-FD failure detection algorithm, and the configurator is informed by the devices that have detected the failure (1).
The configurator retrieves the applications that the failed component was part of from the knowledge base and finds replacement for these devices that are available in the network (2).
Then, the set of applications and available devices is passed to the allocator (3), which computes an allocation for tasks and devices, and returns the resulting allocation to the configurator (4), which then applies the new configuration to devices in the network (5).

\begin{figure}
  \centering
  \begin{tikzpicture}[
    component/.style={draw, thick, align=center, minimum width=1cm, minimum height=1cm},
    every edge/.style={draw, thick, >=Latex[]},
    node distance=0.5cm and 0.8cm,]
    \node[component] (sensors) {Vibration\\sensing};
    \node[component, below right=of sensors] (preprocessing) {Data\\preprocessing};
    \node[component, right=of preprocessing] (kpi) {KPI\\calculation};
    \node[component, right=of kpi] (postprocessing) {Postprocessing};
    \node[component, above=of postprocessing] (fault) {Signal-based\\fault detection};
    \node[component, right=of postprocessing] (decision) {Maintenance\\decision};

    \path[->] (sensors) edge (fault) edge (preprocessing);
    \path[->] (preprocessing) edge (kpi);
    \path[->] (kpi) edge (postprocessing);

    \path[<-] (decision) edge (fault) edge (postprocessing);
  \end{tikzpicture}
  \caption{\label{fig:wirel-vibr-analys} Wireless vibration analysis use case (based on: Krügel et al. \cite{kruegel_rotor_2019})}
\end{figure}
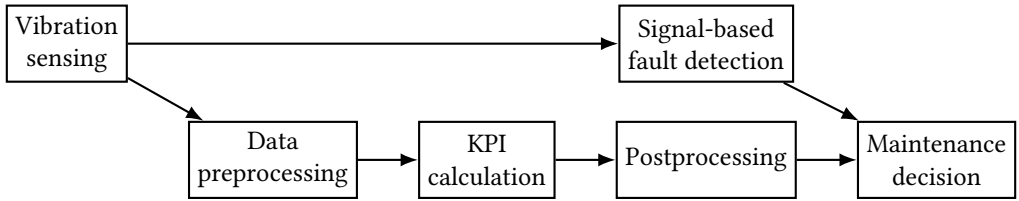

As motivated above, the usage of such a system for optimally self-healing IoT choreographies is increasingly important.
As an example use case for this system, we discuss here the vibration analysis of rotating machinery via vibration sensors as described by Krügel et al. \cite{kruegel_rotor_2019}.
The structure of the application is shown in Figure~\ref{fig:wirel-vibr-analys}.
Vibration sensors sense the vibrations of a rotating machine (such as an engine or fan).
The vibration information is preprocessed for the analysis and fed into a reduced-complexity model of the machine.
From this model, the key performance indicators are derived.
These KPIs are forces acting on the machine parts.
The forces are then postprocessed, and finally, a maintenance decision is made.
In parallel, signal-based fault detection based on flags can detect faults.

It is easy to see that these components of the vibration analysis have different requirements for processing power and required resources.
The sensor nodes running the analysis are battery powered and wireless, since they need to be non-invasive and placed on machines without introducing extra infrastructure.
As such, using an energy-optimal allocation for tasks is important to maximize the runtime of the analysis.

\section{Evaluation}

In this section we first evaluate our failure detector, PE-FD (Section~\ref{sec:evaluation-iota-fd}), as well as the mitigator component for optimal allocation of tasks (Section~\ref{sec:evaluation-mitigator}).

\subsection{Evaluation of PE-FD Algorithm}\label{sec:evaluation-iota-fd}

\begin{figure}
  \centering
  \begin{subfigure}{.5\textwidth}
    \centering
    \includegraphics[width=\textwidth]{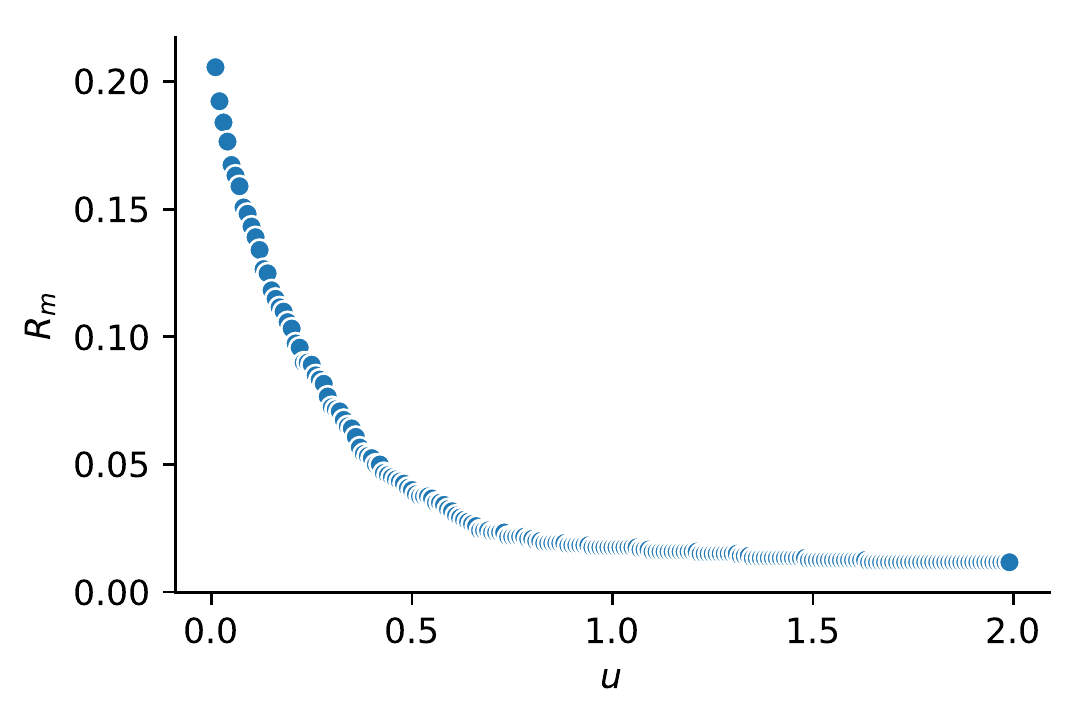}
    \caption{\label{fig:false-positive-rate} False positive rate vs.\ threshold}
  \end{subfigure}%
  \begin{subfigure}{.5\textwidth}
    \centering
    \includegraphics[width=\textwidth]{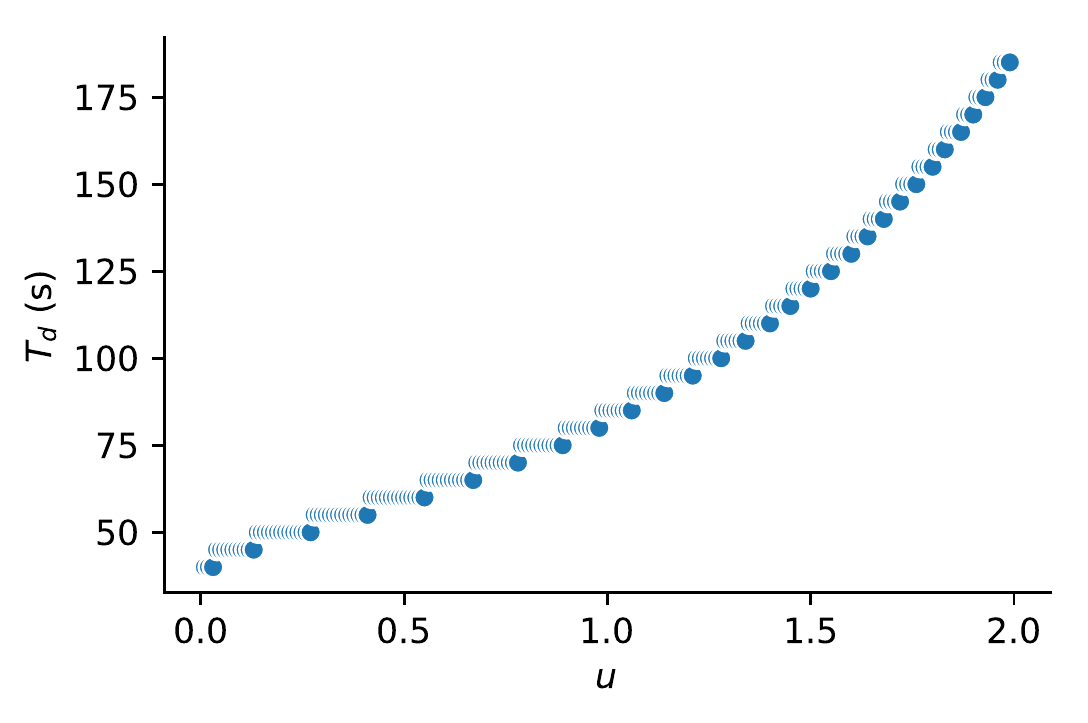}
    \caption{\label{fig:detection-time-s} Detection time (s) vs.\ threshold}
  \end{subfigure}
  \caption{\label{fig:fd-threshold} Behavior of PE-FD with varying thresholds}
\end{figure}

By adjusting the threshold $u$, we can modify the behavior of the failure detection algorithm.
A lower threshold $u$ leads to faster detection of failure, while increasing $u$ reduces the amount of false positives.
Figure~\ref{fig:fd-threshold} shows the behavior of an example failure detection run.
We chose a normally-distributed timestamp arrival time.
We generated timestamp inter-arrival time with a mean of 20 seconds and a variance of 5 seconds for 3000 seconds, and then increased the mean of the distribution to 50 for another 3000 seconds.
This might happen when the sending node switches into an energy saving mode, or changes its network connection to one with increased latency.
We then calculated the detection time (first correct ``failed'' verdict) and mistake rate (incorrect ``failed'' verdicts) for thresholds $u$ from 0.1 to 2.0.
As seen in Figure~\ref{fig:fd-threshold}, increasing the threshold decreases the false positive rate, but decreases the detection time.

Selecting such a threshold can be done with application-level policies based on reconfiguration policies for a node.
A node for which replacements are available can be configured with a lower threshold, since replacing it on a false positive will still allow the system to function, and the node will be replaced faster on a ``true'' failure.

\begin{figure}
  \centering
  \includegraphics[height=.2\textheight]{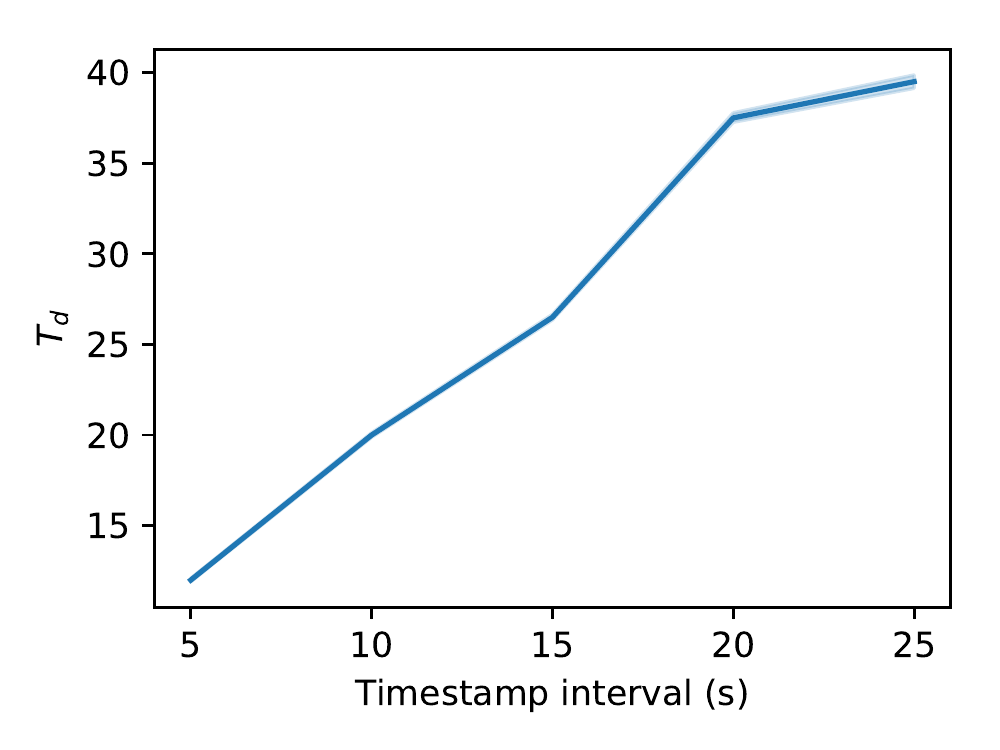}
  \caption{\label{fig:detection-time-vs} Detection time vs.\ timestamp interval.}
\end{figure}

Adjusting the timestamp interval has an impact on the detection time.
By lowering the timestamp interval, the detection time is decreased at an increased cost of network traffic.

Additionally, lowering the timestamp interval can drastically decrease battery life for energy-starved nodes, as waking up and sending packets consumes a large amount of energy.

An example for this behavior in 10 runs of the PE-FD can be seen in Figure~\ref{fig:detection-time-vs}.
We generated timestamp times that were normally distributed around the sending interval with a variance of 1 to account for network delays for a period of 1000 seconds.
We configured PE-FD with a threshold of 0.8, and an infinite learning window.
We then sampled the suspicion function every 5 seconds, and measured the detection time (i.e.\ the first ``true'' detection) for the resulting suspicion values.

We thus see it is advantageous to adjust the timestamp interval of the algorithm dynamically, trading off between the importance of high detection speed (which might be mandated by QoS requirements made by the user) and network and battery efficiency.
Policies to adjust the timestamp interval should take into account the ``kind'' of node they are operating on (wireless or wired network, battery or mains powered) to get achieve optimal results.

\begin{figure}
  \centering
  \begin{subfigure}{.5\linewidth}
      \includegraphics[width=\linewidth]{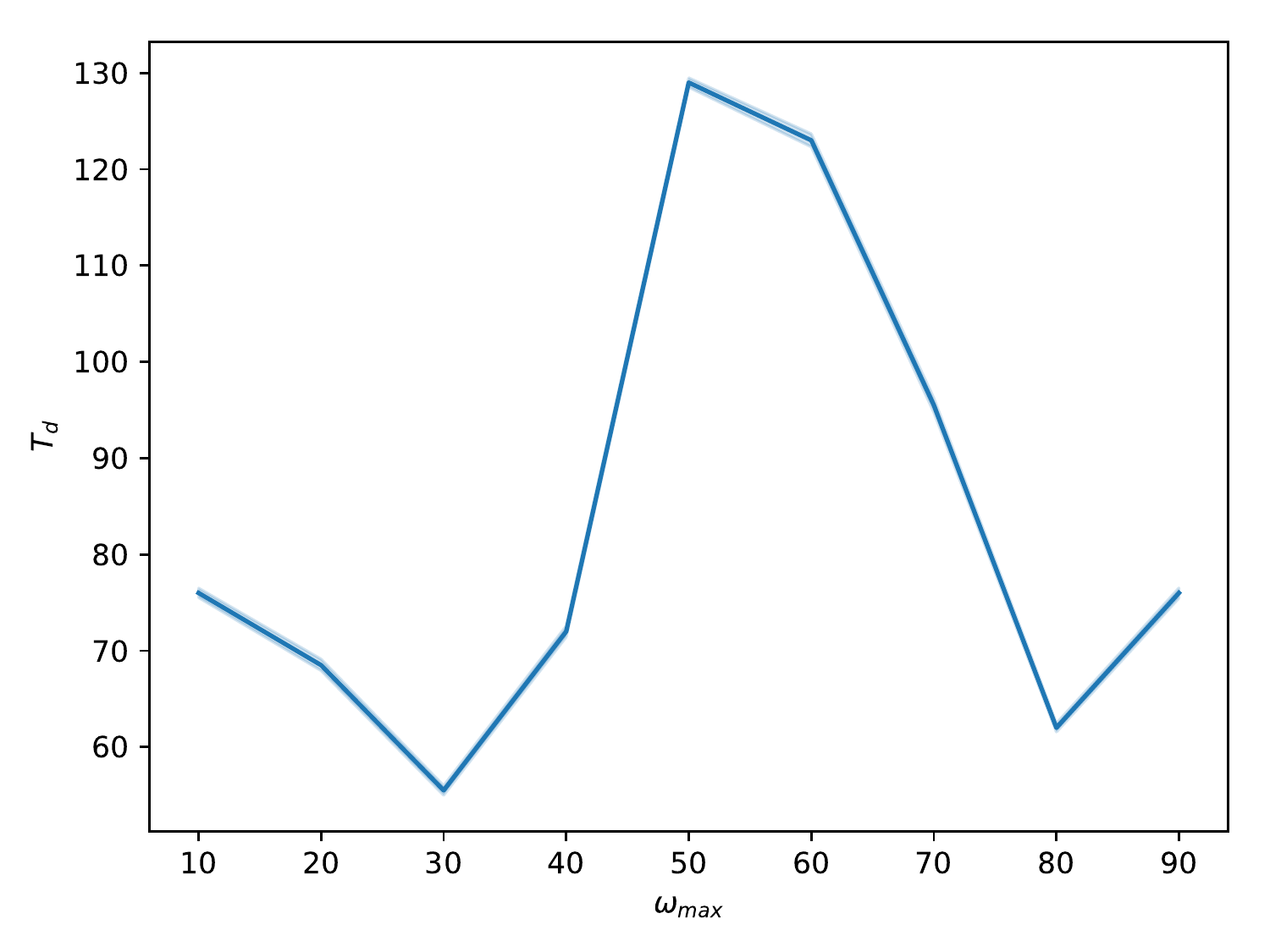}
      \caption{\label{fig:detection-time-vs-1} Detection time vs.\ $\omega_{\text{max}}$.}
    \end{subfigure}%
    \begin{subfigure}{.5\linewidth}
      \includegraphics[width=\linewidth]{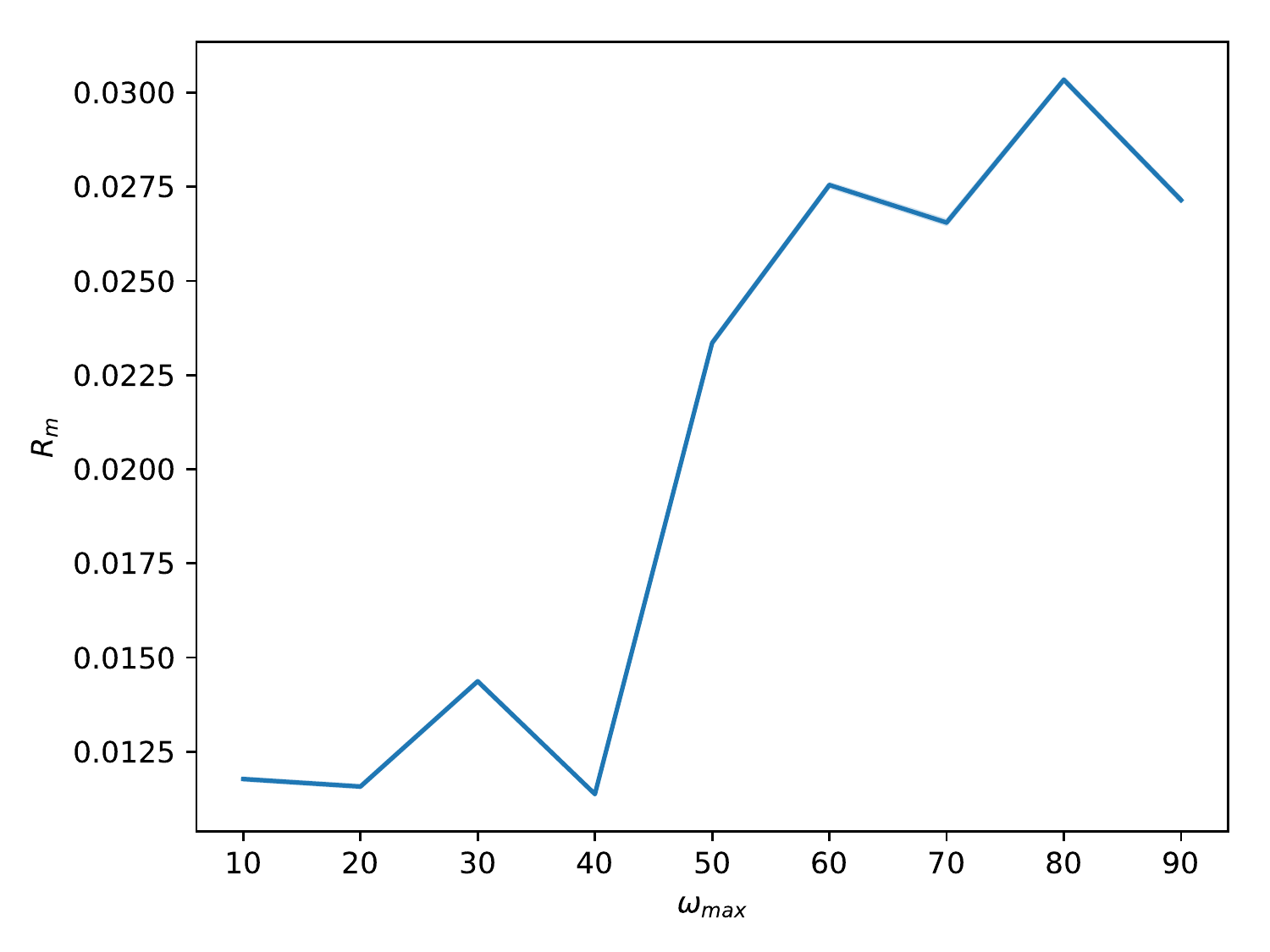}
      \caption{\label{fig:detection-time-vs-2} Mistake rate  vs.\ $\omega_{\text{max}}$.}
    \end{subfigure}
    \caption{\label{fig:omax-effect} Effect of learning window on parameters.}
\end{figure}

The final parameter remaining is the size of the learning window.
We evaluated different configurations of PE-FD with varying window sizes.
The timestamps for this experiment were generated with a normal distribution.
We sampled timestamp inter-arrival times from $\mathcal{N}(20, 1)$ for 1000 seconds, from $\mathcal{N}(50, 1)$ for 1000 seconds, and again from $\mathcal{N}(20, 1)$ for 500 seconds.
This resulted in a total of 95 timestamps.
Figure~\ref{fig:omax-effect} shows the effect of $\omega_{\text{max}}$ on mistake rate and detection time.
The graph shows an interesting trend at an $\omega_{\text{max}}$ of 50.
The timestamp generation generated approximately 50 timestamps (1000 seconds / 20 seconds interval) with a 20 second delay.
This means the learning window of the \omax=50 configuration was reset right as the distribution was changing.
Also, 50 is the largest configuration smaller than the ``period'' of timestamp changes.
This means that this configuration adapts most slowly to changes.
We can see this in the strong growth of both mistake rate and detection time.
Smaller \omax{} configurations adapt faster, while larger \omax{} configurations ``smear'' across the two timestamp distributions and learn a ``mixed'' distribution with a mean between 50 and 20, and a higher variance.
We see this in the graph by the detection time decreasing with larger \omax{}.
The majority of incorrect suspicions are generated at the change from $\mathcal{N}(20, 1)$ to $\mathcal{N}(50, 1)$, at which time there were only 50 timestamps.
Thus, the configurations with \omax{} greater than 50 perform the same as the \omax=50 configurations.
Since the timestamp distribution decreased in average delay, generally, the change from 50 to 20 generated almost no suspicion, as average timestamp times decreased, and thus, the suspicion calculated was set to zero for most samplings of the suspicion function (see Equation~(\ref{eq:15}).).
The general takeaway is that setting the learning window is difficult.
If periodic changes in the timestamp distribution are expected, care should be taken to select a learning window smaller than the period of change.
If no periodic changes are expected, a large learning window should decrease false positives.

\subsection{Evaluation of Mitigator}\label{sec:evaluation-mitigator}

To evaluate the performance of the mitigator, we have built a Python-based evaluation framework.
We evaluate the performance of the mitigator by generating a random network and a random recipe, and letting the allocator find the optimal allocation.

We generate the network with two classes of nodes: Wireless nodes are connected via an energy-inefficient wireless connection, and wired nodes are connected via an energy-efficient wireless connection.
In our configuration, 60\% of the nodes are wired nodes, and the remaining 40\% are wireless nodes.
Nodes are connected to each other with a certain probability.
That probability is 0.8 for wired-wired connections, 0.5 for wireless-wireless connections and 0.4 for wireless-wired connections.
Wired connections use 0.2 units of energy, while wireless connections use 0.8 units of energy.
Nodes have a varying amount of resources uniformly distributed between a lower bound of 1 and an upper bound of 8 resource units.
Nodes also have a varying processing speed between 1 and 3 speedup compared to a reference processor.
Finally, nodes can use from 0.5 to 1.5 as much energy as a reference processor for a single unit of computation.

\begin{figure}
  \centering
  \includegraphics[width=.7\textwidth]{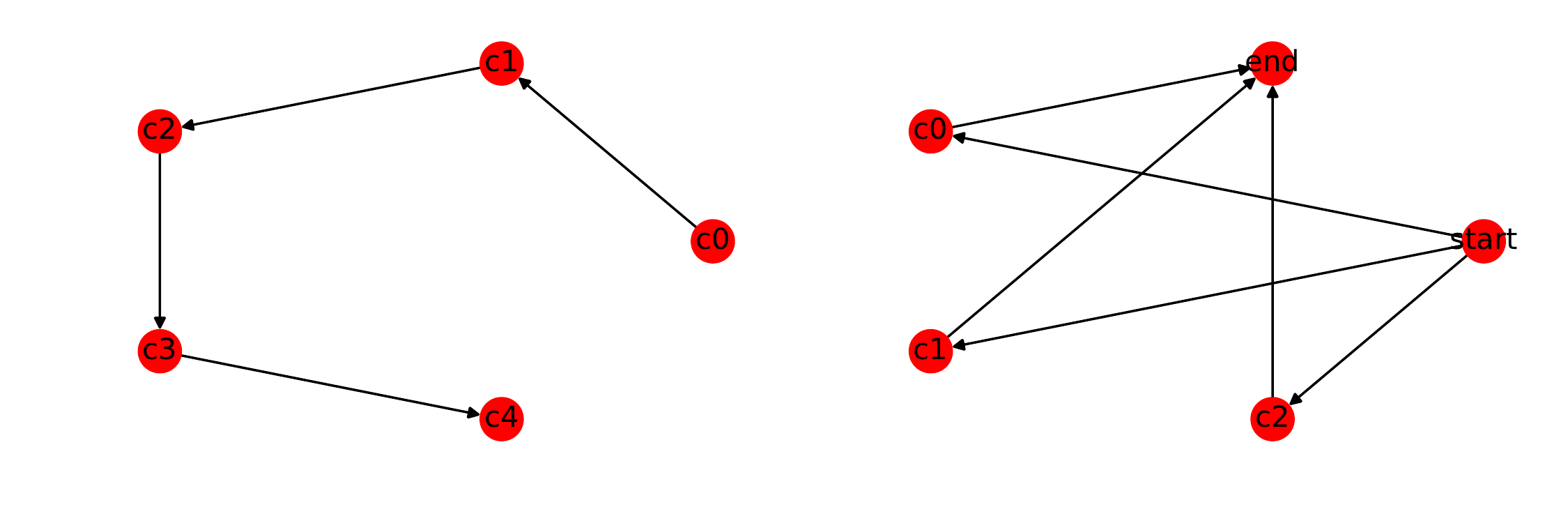}
  \caption{\label{fig:long-wide-recipes} ``Long'' (left) and ``wide'' (right) recipes.}
\end{figure}

For the recipe, we generate two classes of recipes with a certain number of tasks, a ``wide'' recipe and a ``long'' recipe.
In a ``wide'' recipe, two tasks are designated the ``start'' and ``end'' tasks, and every other task needs input from the start node and sends output to the end node.
In a long recipe, tasks are linked serially.
Figure~\ref{fig:long-wide-recipes} shows two example recipes.
Each recipe task has resource requirements randomly distributed between 1 and 8, an output factor randomly distributed between 0.5 and 1.5, and a computation size of 1 or 2.

\begin{figure}
  \centering
  \begin{subfigure}{.45\textwidth}
    \includegraphics[width=.99\textwidth]{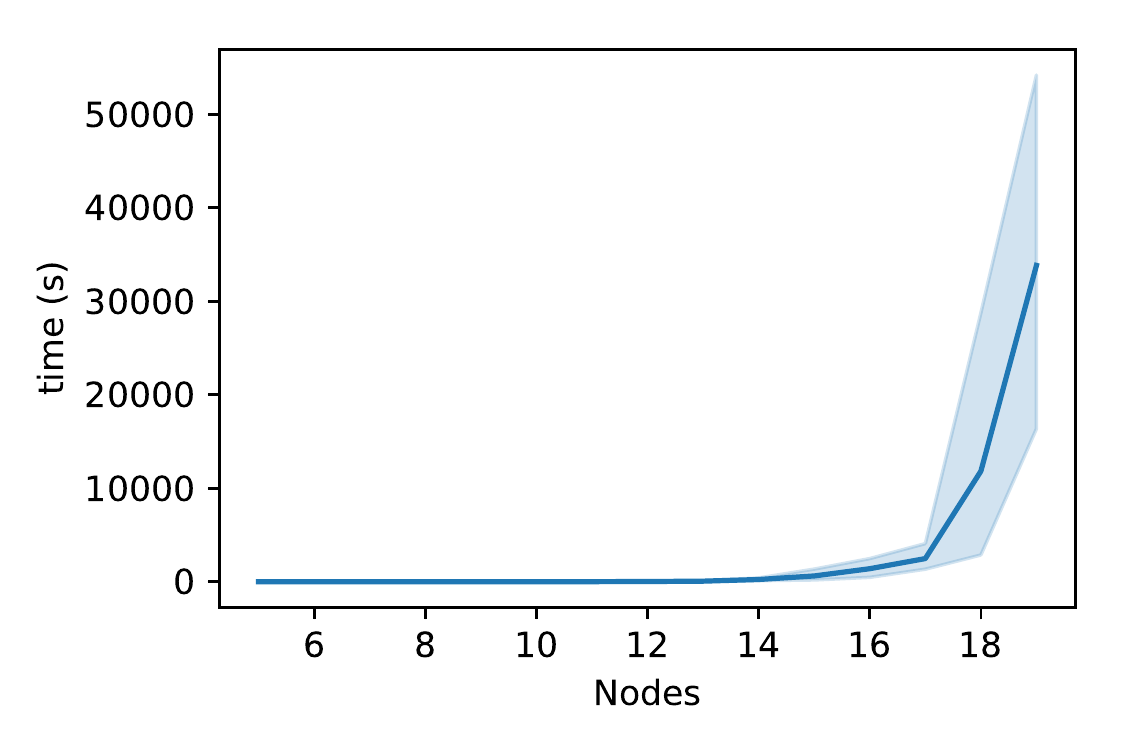}
    \caption{\label{fig:runt-optim-alloc-nodes} CPU time for the optimal allocation algorithm vs.\ the number of nodes.
      Each experiment with n nodes was measured 5 times with 3 to n-1 tasks.}
  \end{subfigure}%
  \begin{subfigure}{.45\textwidth}
      \includegraphics[width=.99\textwidth]{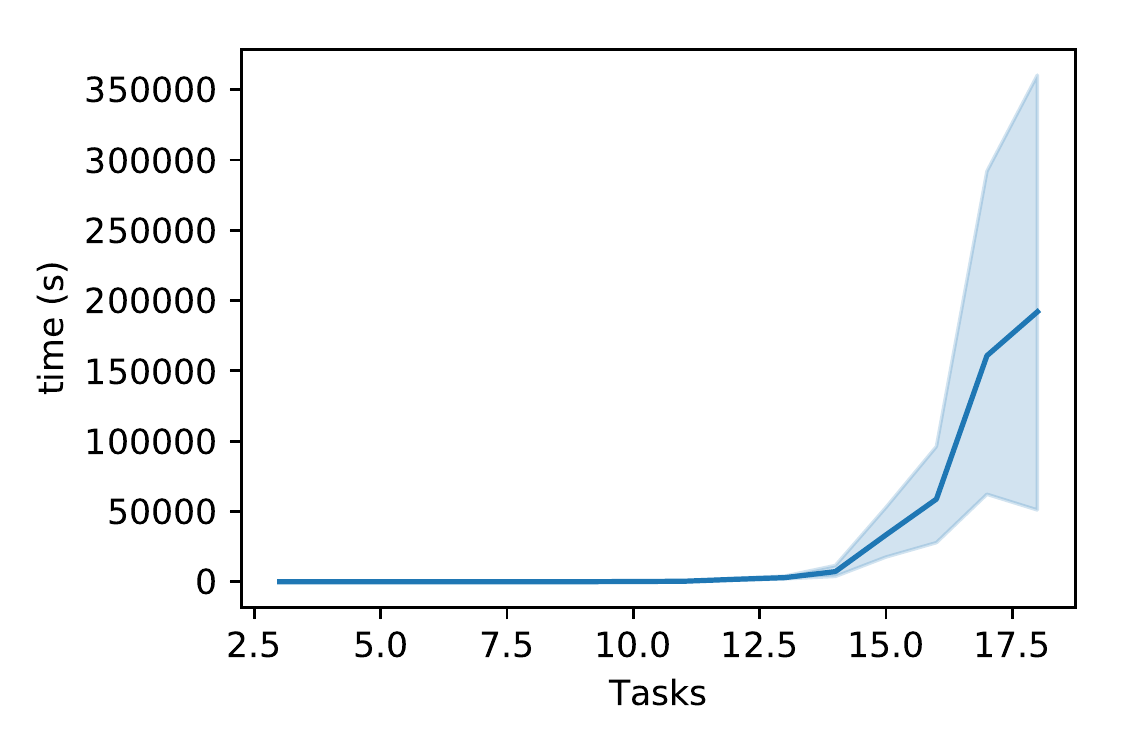}
      \caption{\label{fig:runt-optim-alloc-tasks} CPU time for the optimal allocation algorithm vs.\ the number of tasks.
        Each experiment with n tasks was measured 5 times with 5 to 20 nodes.}
    \end{subfigure}
    \caption{\label{fig:runt-optim-alloc} Runtime for optimal allocation.}
\end{figure}

As expected, the optimal allocation algorithm scales very badly (non-polynomially).
In Figure~\ref{fig:runt-optim-alloc}, we see the runtime of the algorithm for varying problem sizes.
The shaded area shows the variance with the non-shown parameter (different recipe sizes for the network node graph, differing network sizes for the recipe node graph).
The time needed for finding the optimal allocation grows unwieldy very quickly.

\begin{figure}
  \centering
  \begin{subfigure}{.45\textwidth}
    \centering
    \includegraphics[width=\textwidth]{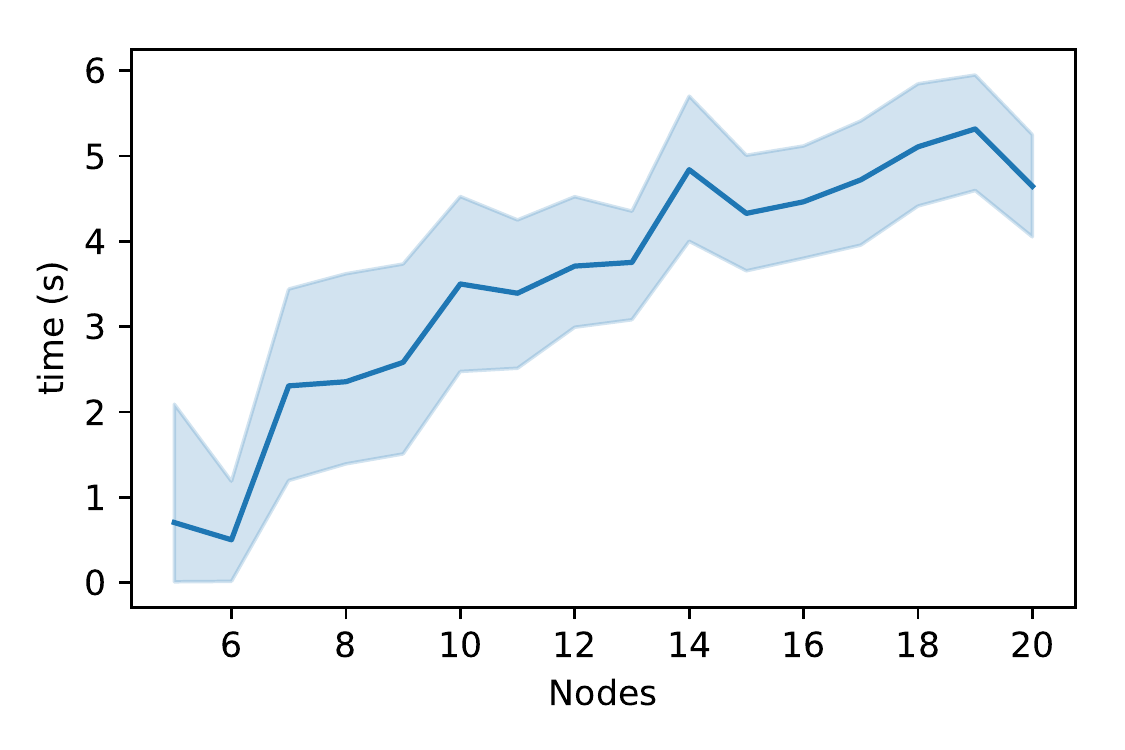}
    \caption{\label{fig:runt-alloc-heur-nodes} CPU time of the allocation heuristic vs.\ the number of nodes. Each experiment with n nodes was measured 5 times with 3 to n-1 tasks.}
  \end{subfigure}%
  \begin{subfigure}{.45\textwidth}
    \centering
    \includegraphics[width=.99\textwidth]{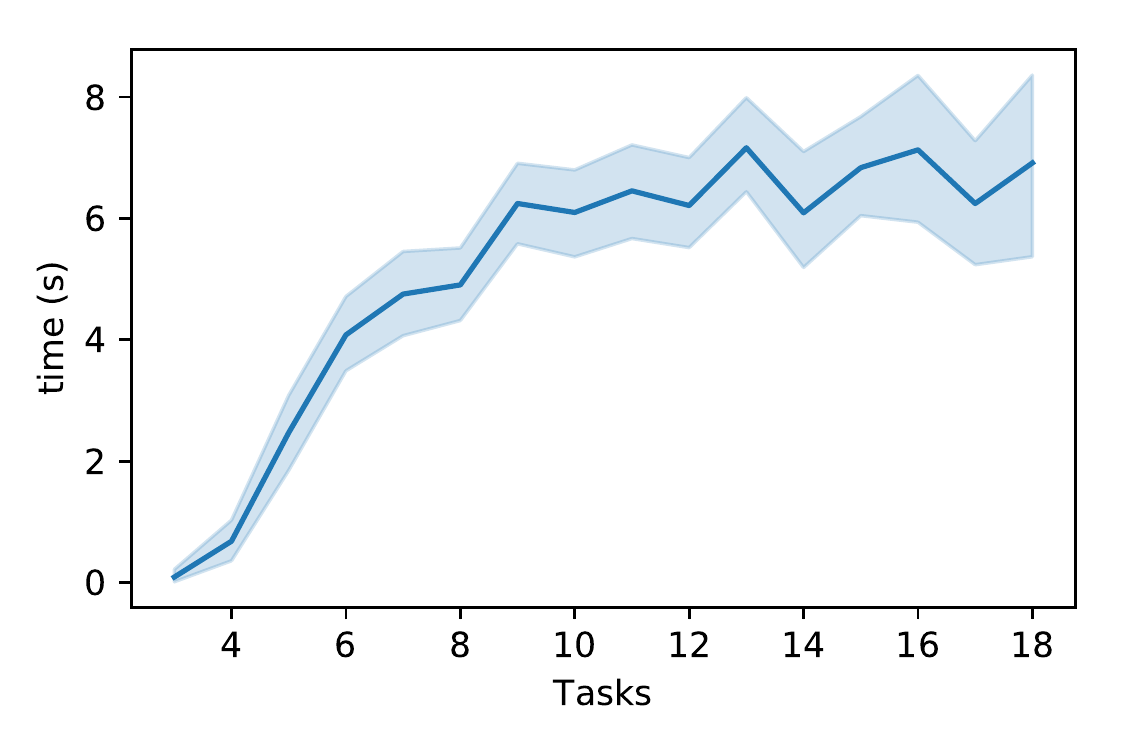}
    \caption{\label{fig:runt-alloc-heur-tasks} CPU time of the allocation heuristic vs.\ the number of tasks. Each experiment with n tasks was measured 5 times with 5 to 20 nodes.}
  \end{subfigure}
  \caption{\label{fig:runt-alloc-heur} Heuristic runtime}
\end{figure}

In comparison, our heuristic presented in Section~\ref{sec:optim-energy-optim} finds a solution much more quickly.
Figure~\ref{fig:runt-alloc-heur} shows the runtime of the heuristic for different network and recipe sizes.
For the slowest case for the full allocation, the heuristic takes 8 seconds of CPU time, while the solver consumes 864104 seconds (about 10 days) of CPU time for finding the optimal allocation.
The allocation evaluation was executed on an Amazon EC2 \texttt{m4.10xlarge} machine with 40 virtual cores and 160 GiB of memory.
Peak memory use was 51 GiB\@.

\begin{figure}
  \centering
  \includegraphics[height=.3\textheight]{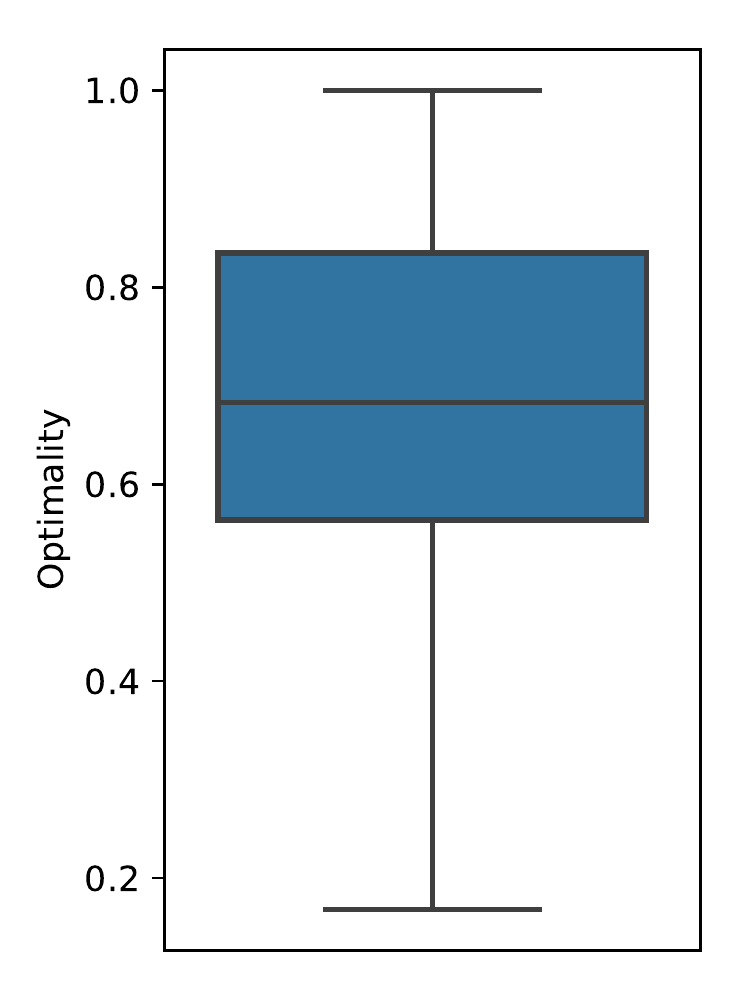}
  \caption{\label{fig:energy-cons-heur} Energy consumption of heuristic solution scaled against optimal solution.}
\end{figure}

However, the heuristic loses about 30\% of energy efficiency over the optimal algorithm.
As seen in Figure~\ref{fig:energy-cons-heur}, 50\% of the solution lie in the 0.6 to 0.8 range.

\begin{figure}
  \centering
  \includegraphics[height=.2\textheight]{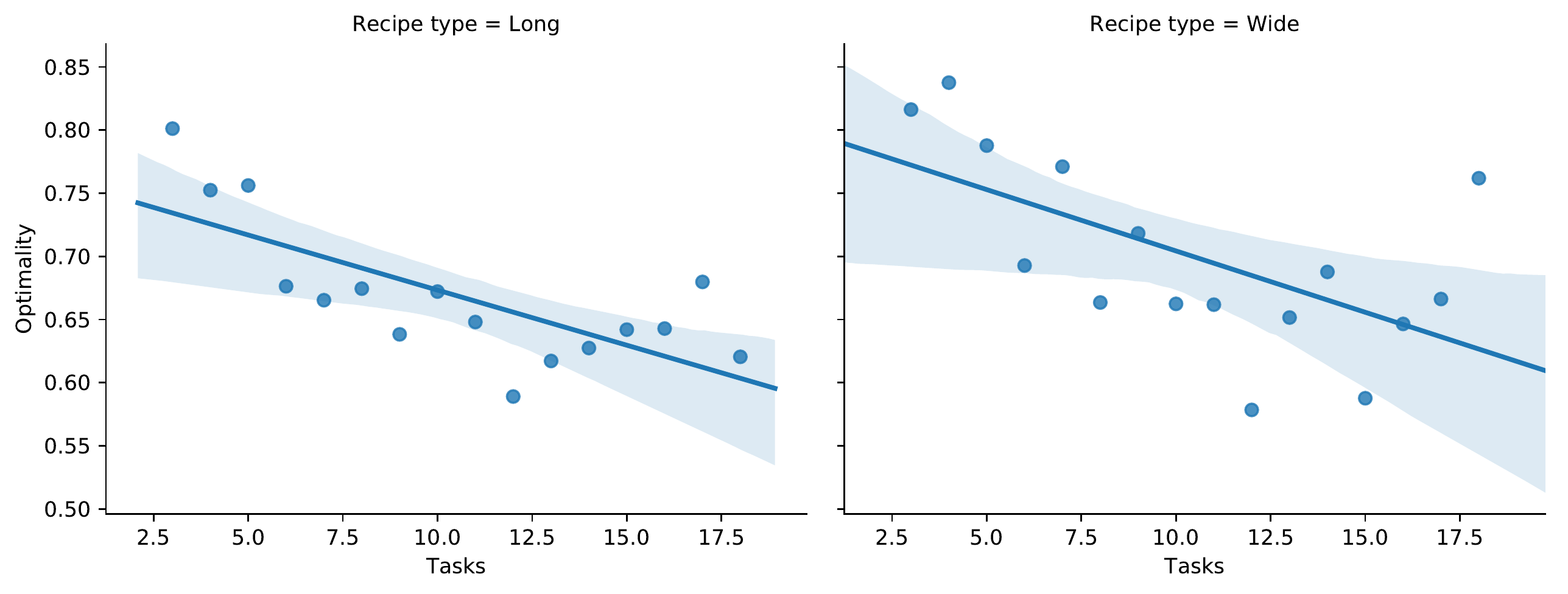}
  \caption{\label{fig:perf-heur-vs} Performance of heuristic vs.\ number of tasks and recipe type.}
\end{figure}

In Figure~\ref{fig:perf-heur-vs}, the performance of the heuristic as related to the size of the recipe can be see.
The performance of the heuristic decreases with larger networks.
This is explainable by the network links longer, as the difference between the node-local transmission energy $\hat{T}_n$ and the actual transmission energy $T_{(n, n2)}$ grows larger with a growing network.
The diagram also shows that the long recipe is harder to allocate than the wide recipe.

This improves the results shown in~\cite{cardellini16optimal_operator_placement} for the sampling heuristic, where a sampling factor of 30\% (i.e.\ only one third of all nodes was considered) led to a runtime of 5 seconds, but solution quality of 40\% for the sequential application.

\section{Conclusions \& Future Work}\label{sec:conclusions--future}

Today, IoT applications are increasingly executed in edge environments to avoid latency and privacy issues associated with a cloud-based execution. To enable the execution of complex applications on the edge, we need to split them in separate tasks and execute them on multiple devices. An example of such a complex application has been described above: the vibration analysis of rotating machinery on the manufacturing shop floor. Running such distributed applications reliably is a challenge.

We present in this work a system that supports the self-healing of such IoT choreographies. This system consists mainly of two contributions:
(1) A novel failure detector concept that supports a wide range of parameters for application-specific and policy-based configuration and some guidelines towards the selection of these parameters.
(2) We have introduced an ILP formulation for optimal task allocation with regards to energy, and designed a heuristic that makes on-line computation of allocations feasible.
We have evaluated both the PE-FD failure detector and the performance of the allocation algorithm.

In the future, we plan to formalize the policies described textually in Section~\ref{sec:polic-appl-tuned}, possibly in the form of semantic rules in combination with a reasoner.
In conjunction with formally described application requirements, this will allow to automatically infer a tuned failure detector parameterization without manual configuration.

Further, we aim to extend our allocation benchmarking framework to evaluate other heuristics for allocation, taking into account resource distribution in the network.
These heuristics will likely perform better on large networks, where the simple ``outgoing'' heuristic fails.

Additionally, realizing and evaluating the use case as described in Section~\ref{sec:system-model} will be required to gain a better understanding of the use of allocation and failure detection in industrial automation systems.
Also, we aim to integrate our system with the Node-RED framework for the easy  creation of applications.
With approaches such as Distributed Node-RED\cite{giang_developing_2015} and the traction Node-RED is gaining in automation communities, this will be a promising field to apply the techniques described in this paper.

\begin{acks}
  This work is part of the SEMIoTICS project\footnote{\url{https://www.semiotics-project.eu/}} that develops a pattern-driven framework to guarantee secure and dependable behavior in IoT environments~\cite{Fysarakis2019}.
  It received funding from the European Union's Horizon 2020 research and innovation program under grant agreement No.
  780315.
\end{acks}

\bibliographystyle{ACM-Reference-Format}
\bibliography{bibliography}

\end{document}